\documentclass{aastex631}

\usepackage{array}
\usepackage{tikz}
\usetikzlibrary{shapes.geometric, arrows}
\usetikzlibrary{positioning}

\tikzset{
    startstop/.style={rectangle, rounded corners, minimum width=2.6cm,
        minimum height=1cm, text centered, draw=black, fill=gray!20},
    process/.style={rectangle, minimum width=3.2cm, minimum height=1cm,
        text centered, draw=black, fill=blue!20},
    decision/.style={diamond, aspect=2, text centered,
        draw=black, fill=green!20, inner sep=1pt},
    arrow/.style={thick,->,>=stealth}
}
\shorttitle{Energy Functions and Cosmic Evolution of FRBs}
\shortauthors{Dong et al.}

\begin{document}

\title{Energy Functions and Cosmic Evolution of Repeating and Non-repeating Fast Radio Bursts}

\correspondingauthor{Yong-Feng Huang}
\email{hyf@nju.edu.cn}
\correspondingauthor{Zhi-Bin Zhang}
\email{z\_b\_zhang@sina.com}

\author[0009-0000-0467-0050]{Xiao-Fei Dong}
\affiliation{School of Astronomy and Space Science, Nanjing
University, Nanjing, 210023, China}
\author{Qing Pan}
\affiliation{College of Physics and Physical Engineering, Qufu
Normal University, Qufu, 273165, China}
\author[0000-0001-7199-2906]{Yong-Feng Huang}
\affiliation{School of Astronomy and Space Science, Nanjing
University, Nanjing, 210023, China}
\affiliation{Key Laboratory of
Modern Astronomy and Astrophysics (Nanjing University), Ministry
of Education, China}
\author[0000-0003-4859-682X]{Zhi-Bin Zhang}
\affiliation{College of Physics and Physical Engineering, Qufu
Normal University, Qufu, 273165, China}
\author[0000-0001-7892-9790]{Hao-Xuan Gao}
\affiliation{School of Physics and Astromoy, Anqing Normal
University, Anqing 246133, China}
\affiliation{Institute of
Astronomy and Astrophysics, Anqing Normal University, Anqing
246133, China}
\author[0000-0001-9648-7295]{Jin-Jun Geng}
\affiliation{Purple Mountain Observatory, Chinese Academy of
Sciences, Nanjing, 210023, China}
\author[0000-0002-6299-1263]{Xue-Feng Wu}
\affiliation{Purple Mountain Observatory, Chinese Academy of
Sciences, Nanjing, 210023, China}
\affiliation{School of Astronomy
and Space Sciences, University of Science and Technology of China,
Hefei, 230026, China}
\author[0000-0003-0721-5509]{Lang Cui}
\affiliation{State Key Laboratory of Radio Astronomy and
Technology, Xinjiang Astronomical Observatory, CAS, 150 Science
1-Street, Urumqi 830011, China}
\affiliation{Xinjiang Key
Laboratory of Radio Astrophysics, 150 Science 1-Street, Urumqi,
Xinjiang, 830011, China}
\author[0009-0009-1579-5209]{Xiu-Juan Li}
\affiliation{School of Cyber Science and Engineering, Qufu Normal
University, Qufu, 273165, China}
\author[0000-0003-2366-219X]{Song-Bo Zhang}
\affiliation{Purple Mountain Observatory, Chinese Academy of
Sciences, Nanjing, 210023, China}
\author[0000-0002-2162-0378]{Abdusattar Kurban}
\affiliation{State Key Laboratory of Radio Astronomy and
Technology, Xinjiang Astronomical Observatory, CAS, 150 Science
1-Street, Urumqi 830011, China}
\affiliation{Xinjiang Key
Laboratory of Radio Astrophysics, 150 Science 1-Street, Urumqi,
Xinjiang, 830011, China}
\author[0000-0002-5238-8997]{Chen-Ran Hu}
\affiliation{School of Astronomy and Space Science, Nanjing
University, Nanjing, 210023, China}
\author[0000-0002-2191-7286]{Chen Deng}
\affiliation{School of Astronomy and Space Science, Nanjing
University, Nanjing, 210023, China}
\affiliation{Key Laboratory of
Modern Astronomy and Astrophysics (Nanjing University), Ministry
of Education, China}

\begin{abstract}

Fast radio bursts (FRBs) are millisecond radio pulses of unknown
origin. Despite extensive follow-up observations, only $\sim3\%$
of FRBs have been confirmed as repeaters. It remains unclear
whether the rest are truly one-off bursts, or essentially
repeating sources that have only been detected once due to limited
monitoring time. Using the second CHIME/FRB catalog, we test this
debate by comparing non-repeaters with two repeater-based
subsamples: the first-detected bursts of repeaters and their
highest-fluence bursts. A non-parametric method that accounts for
selection effects is employed to derive the energy functions and
event rates of these samples. All samples are well described by
broken power-law energy distributions with comparable break
energies ($\sim 5\times10^{38}$ erg), but with significantly
different slopes between repeating and non-repeating populations.
Their event-rate evolution also differ significantly. Assuming
$\rho(z) \propto (1+z)^B$, we have $B = -5.57^{+0.15}_{-0.15}$ for
non-repeaters and $B = -8.63^{+0.46}_{-0.41}$ and
$-9.10^{+0.55}_{-0.56}$ for the two repeater samples. The
differences in cosmic event-rate evolution between non-repeating
and repeating FRBs persist across the tested assumptions,
suggesting possible population-level differences between repeating
and apparently non-repeating FRBs.

\end{abstract}


\section{Introduction}

Fast radio bursts (FRBs) are mysterious bright,
millisecond radio pulses
originating from cosmological distances
\citep{2007Sci...318..777L, 2020Natur.587...54C, 2026ApJS..283...34C}.
Following the discovery of repeated bursts from FRB 20121102,
the FRB population was subsequently classified into
repeaters and non-repeaters \citep{2016Natur.531..202S}.
Yet, as of April 2026, only 101 FRBs have been confirmed
as repeaters, while over $97\%$ of detected sources remain one-off
events \citep{2023Univ....9..330X}. Although only a small fraction of FRBs have been observed to
repeat, this does not rule out the possibility that all FRBs are
intrinsically repeaters \citep{2019ARA&A..57..417C,
2024MNRAS.52711158Y, 2025arXiv251019143K, 2026arXiv260508410C}.

The fact that ``One-off'' events were detected only once could be
due to limited sensitivity or monitoring time, which means they
may still be capable of repeating \citep{2019NatAs...3..928R,
2020MNRAS.495.2416J, 2024Ap&SS.369...59L, 2024NatAs...8..337K,
2025ApJ...993...37B}. Observational evidence also suggests similarities between the two classes, with no statistically significant differences established in their host-galaxy properties \citep{2024Natur.635...61S}. Moreover, repeating FRBs can also occur in old stellar environments, as exemplified by FRB 20200120E in a globular cluster \citep{2021ApJ...910L..18B, 2025arXiv250801648C}.

Nevertheless, repeating and apparently non-repeating FRBs exhibit several observational differences. Repeating FRBs tend to exhibit broader pulse widths, narrower bandwidths, lower extragalactic dispersion measures (DMs), and lower characteristic energies than apparently non-repeating FRBs \citep{2021ApJS..257...59C, 2021ApJ...923..230L, 2022Univ....8..355Z, 2023Univ....9..251Z, 2026ApJ...998..154K}. Some repeaters show large rotation measures (RMs) and strong depolarization \citep{2022Sci...375.1266F}.
Considerable diversity also exists among repeating FRBs in their DM and RM evolution, polarization properties, pulse morphology, and periodic activity \citep{2019ApJ...876L..23H, 2020Natur.586..693L, 2021ApJ...917...13S, 2023ApJS..269...17H, 2025arXiv250312013D}.

The energy functions and cosmic event-rate evolution of FRBs provide important statistical constraints on their populations. Previous studies have described FRB energy distributions using power-law \citep{2019ApJ...883...40L}, broken power-law \citep{2023ApJ...955..142Z}, and Schechter-like functions \citep{2022MNRAS.511.1961H}. \citet{2024NatAs...8..337K} found that the high-energy distribution of an active repeater FRB 20220912A resembles that of apparently non-repeating FRBs.
Studies of the Canadian Hydrogen Intensity Mapping Experiment (CHIME; \citealt{2018ApJ...863...48C}) FRB Catalog 1 found declining cosmic event rates of apparently non-repeating FRBs using the $V_{\max}$ method \citep{2022MNRAS.511.1961H} and non-parametric methods \citep{2024ApJ...973L..54C}, with the latter yielding an evolution index of $-4.9\pm0.3$. Similar declines were reported by \citet{2025ApJ...988L..64C} and \citet{2025A&A...698A..18Z}. However, studies accounting for more detailed selection effects found weaker constraints, with star formation rate (SFR)-tracing models remaining viable \citep{2023ApJ...944..105S, 2025A&A...698A.127M}. More recently, a non-parametric study of CHIME/FRB Catalog 2 found a rate evolution index of $-5.38\pm0.02$ \citep{2026ApJ..1003..179J}.

Despite these efforts, a systematic comparison of the energy functions and cosmic event-rate evolution of repeating and apparently non-repeating FRBs remains lacking. Here, we use CHIME/FRB Catalog 2 to compare these properties within a unified non-parametric framework, considering both the first-detected and the highest-fluence bursts of identified repeaters.

This paper is organized as follows. In Section~\ref{sec: DATA},
detailed data acquisition are described and three FRB samples used
for the analysis are introduced. Section~\ref{sec: METHOD}
describes the nonparametric method. The results are presented in
Section~\ref{sec: RESULTS}. Section~\ref{sec:test} presents robustness tests of our results. Our conclusions and a brief discussion
are presented in Section~\ref{sec: Conclusion and Discussion}.

\section{DATA}
\label{sec: DATA}

More than 4000 FRB
sources\footnote{https://blinkverse.zero2x.org/} have been
detected by different radio facilities. To minimize
survey-dependent selection effects, we restrict our analysis to
FRBs detected by CHIME \citep{2018ApJ...863...48C}). The data are
taken from the second CHIME/FRB Catalog
\footnote{https://www.chime-frb.ca/catalog2}
\citep{2026ApJS..283...34C} and archived at the Canadian Advanced Network for Astronomical Research \citep{CHIME2025data}, from which we select those bursts having a
signal-to-noise ratio greater than 10, with reliable DM and
measured fluence (${\cal F_\nu}$). The resulting sample contains
3057 non-repeaters and 78 repeaters. Below, we describe the
redshift estimation and the construction of the subsamples used in
this study.

\subsection{Redshift and energy}

The redshift is not measured for most FRBs. We thus estimate this
parameter from their DM value. The observed ${\rm DM}$ of an FRB
includes several components
\begin{equation}\label{DM}
{\rm DM_{obs}} = {\rm DM_{MW}} + { \rm DM_{IGM}} + {\rm
\frac{DM_{host} + DM_{src}}{(1+z)}},
\end{equation}
where ${\rm DM_{MW}}$, ${\rm DM_{IGM}}$, ${\rm {DM_{host}}}$ and
${\rm DM_{src}}$ are the DM contributions from the Milky Way,
intergalactic medium (IGM), FRB host galaxy and local FRB
environment, respectively \citep{2018ApJ...867L..21Z}. ${\rm
DM_{IGM}}$ is related to the redshift of the source through
\begin{equation}
{\rm DM_{IGM}} =  \frac{3c H_{\rm 0} \Omega_{\rm b} f_{\rm IGM}
}{8 \pi G m_{\rm p}} \times \int_0^z \frac{\chi(z) (1+z) dz}
{[\Omega_{\rm m} (1+z)^3+1-\Omega_{\rm m}]^{1/2}},
 \label{DM_IGM}
\end{equation}
where $H_{\rm 0}$ is the Hubble constant, $\Omega_{\rm b}$ is the
baryon density, $f_{\rm IGM}$ is the fraction of baryons in the
IGM, $m_{\rm p}$ is the mass of proton, and $\chi(z)$ denotes the
free electron number per baryon in the universe
\citep{2014ApJ...783L..35D, 2020ApJ...895...33W}. In this study,
we adopt a flat $\Lambda$CDM cosmology with the Hubble parameter
and the density parameters taken as $H_{\rm 0}=69.6~{\rm
km~s^{-1}~Mpc^{-1}}$ and ${\Omega}_{\rm m}=0.286$
\citep{2014ApJ...794..135B}. Following \cite{2018ApJ...867L..21Z},
we estimate the FRB redshift as
\begin{equation}\label{z}
z \sim {\rm DM_{IGM}} \ /\ 855 \ {\rm pc\ cm^{-3}},
\end{equation}
which is approximate as long as $z<3$.

In order to obtain ${\rm DM_{IGM}}$ of an FRB, one has to
determine ${\rm {DM_{host}}}$, ${\rm DM_{MW}}$ and ${\rm
DM_{src}}$, respectively. It is not straightforward to derive
these quantities from an individual FRB, but may be estimated
statistically from the FRB population. Various assumptions have
been taking in estimating the redshift of FRBs, such as
considering a log-normal distribution of ${\rm {DM_{host}}}$
\citep{2020ApJ...900..170Z}, considering $\rm He\ II$ reionization
\citep{2024ApJ...975..184W} or considering correlations between
the excess dispersion measure and host galaxy properties
\citep{2025arXiv250701270L}.

Following the statistical work of
\cite{2023ApJ...944..105S}, ${\rm {DM_{host}}}$ and ${\rm
DM_{MW}}$ are adopted as $84^{+69}_{-49}\ {\rm pc\ cm^{-3}}$ and
$\sim80{\ \rm pc\ cm^{-3}}$ respectively, as representative values for the CHIME/FRB population. The impact of these assumptions is assessed in Section~\ref{DMtest}.
In addition, the source component ${\rm DM_{src}}$
depends on the FRB progenitor models and is relatively small
\citep{2018ApJ...867L..21Z}. To balance different models, we
assume that the effect of ${\rm DM_{src}}$ is contained in ${\rm
{DM_{host}}}$, i.e., ${\rm {DM_{host}} + {\rm DM_{src}}} \sim 84\
{\rm pc\ cm^{-3}}$.

With redshift derived by Equation~(\ref{z}), one can calculate the
isotropic energy of an FRB  following \cite{2018ApJ...867L..21Z}
as
\begin{equation}
E_{\rm iso} \simeq \frac{4\pi D^2_{\rm L}}{(1+z)} {\cal F_\nu} \nu_c
= (10^{39} \ {\rm erg}) \frac{4\pi}{(1+z)} \left(\frac{D_{\rm L}}{10^{28} \ {\rm cm}}\right)^2 \frac{\cal F_\nu}{\rm Jy \cdot ms} \frac{\nu_c}{\rm GHz},
\label{eq:E}
\end{equation}
where $D_{\rm L}$ is the luminosity distance
related to redshift $z$ as
{\small\begin{equation}
D_{\rm L}(z)=\frac{c}{H_{\rm 0}}(1+z)\int_0^z
\frac{dz}{\sqrt{1-\Omega_{\rm m}+\Omega_{\rm m}(1+z)^3}},
\end{equation}}
and ${\cal F_\nu}$ is the specific fluence (in units of $\rm Jy
\cdot ms$). The central frequency $\nu_c$ is chosen for estimating
the burst energy.

\subsection{Definition of samples}

The 3057 non-repeating FRBs are defined as $S_{\rm NR}$.
For repeaters, each repeating FRB can produce multiple bursts,
and the observed burst properties
may vary from burst to burst.
We therefore select one representative burst from each of the 78 repeating FRBs
to construct the repeater samples.
The first-detected burst from each repeater is used to construct $S_{\rm first}$ sample.
The highest-fluence burst from each repeater is used to construct $S_{\rm max}$ sample.
This operation allows us to investigate the relationships among bursts from
repeating FRBs and the differences between repeaters and apparent one-off events.

\begin{figure*}
     \centering
     \includegraphics[width=1.0\textwidth]{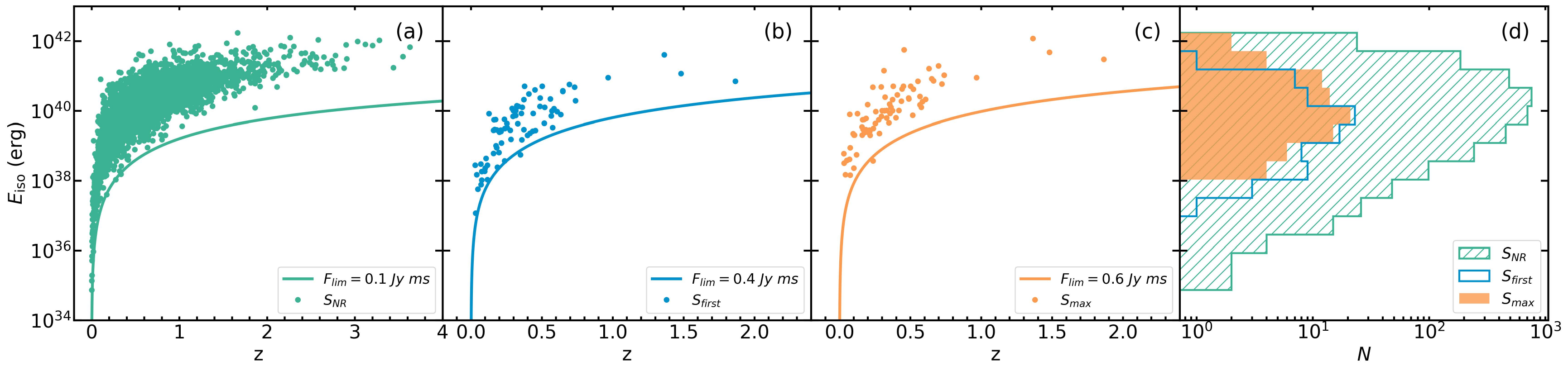}
     \caption{Distribution of the FRB energy and redshift.
     The scatter dots in the panel (a), (b), (c) represent
     $S_{\rm NR}$, $S_{\rm first}$ and $S_{\rm max}$ respectively. The solid lines represent the detection
     threshold of each sample.
     Panel (d) represents the energy distribution of the three samples.}
     \label{fig1}
\end{figure*}

The isotropic energy and estimated redshift distributions of
the three FRB samples are shown in Figure~\ref{fig1}.
Approximately 99\% of the FRBs lie within the redshift range
$0 < z < 2.5$, with only 24 non-repeaters exceeding
this range. The maximum estimated redshift for repeaters is 1.86,
while that for non-repeaters reaches 3.63.
The energy distribution, shown in panel (d) of Figure~\ref{fig1},
spans from $\sim10^{35}$ to $10^{42}\ \rm erg$,
consistent with previous studies
\citep{2021ApJS..257...59C, 2021MNRAS.501..157Z,
2022ApJ...924L..14Z, 2022MNRAS.511.1961H, 2023ApJS..269...17H}.

The solid lines in the first three panels of Figure~\ref{fig1}
correspond to the truncation boundaries, representing the maximum
redshift at which CHIME can detect an FRB of a given energy. The
fluence threshold ($F_{\rm lim}$) are defined as the minimum
detectable fluence of each FRB sample, which are $0.1\ \rm Jy
\cdot ms$, $0.4\ \rm Jy \cdot ms$ and $0.6\ \rm Jy \cdot ms$,
respectively. The detection threshold of energy is calculated as
\begin{equation}
E_{\rm lim} = \frac{4\pi D^2_{\rm L}}{(1+z)} {F_{\rm lim}} {\nu_c}.
\end{equation}

\section{The nonparametric method}
\label{sec: METHOD}
The energy function and event rate of the FRB
population offer key insights into their progenitors across
different evolutionary stages \citep{2019NatAs...3..928R,
2022MNRAS.511.1961H}. The distribution of FRBs is a joint function
of both energy and redshift, which can be written as $\Psi(E, z)$.
If the dependence of the distribution on the two parameters is
independent of each other, then we have $\Psi(E, z) =
\psi(E)\phi(z)$, where $\psi(E)$ represents the energy function
and $\phi(z)$ represents the redshift distribution of FRBs.
However, an apparent correlation between $E$ and $z$ is observed in the first three panels of Figure~\ref{fig1}, which may be affected by observational selection effects.
This observational bias must be taken into account
when reconstructing the intrinsic FRB distributions
from the observational data
\citep{2019A&A...625A.109L, 2020MNRAS.498.3927H,
2022MNRAS.516.4862J}.

The Lynden-Bell's $C^{-}$ method \citep{1971MNRAS.155...95L} is
suitable for dealing with such mutually independent bivariate
problems \citep{Woodroofe1985, Wang1986}, which can help obtain
the event rate of FRBs \citep{2025A&A...698A..18Z,
2025ApJ...988L..64C, 2026ApJ..1003..179J}. As a statistical
method, it has been applied in many other fields of astronomy,
such as short/long gamma-ray bursts \citep{2015ApJS..218...13Y,
2016A&A...587A..40P, 2022MNRAS.513.1078D, 2023ApJ...958...37D,
2024MNRAS.535.2800L, 2025ApJ...993...20D, 2026ApJ..1003..227D},
binary black holes \citep{2024ApJ...977...29D}, galaxies
\citep{1978AJ.....83.1549K, 1986ApJ...307L...1L,
1986MNRAS.221..233P}, pulsars \citep{2016Ap&SS.361..138D},
and quasars \citep{2011ApJ...743..104S,
2021ApJ...913..120Z}, to remove the evolution of parameters
\citep{2013ApJ...774..157D, 2017A&A...600A..98D,
2021ApJ...920..135X}.

The independence of the truncated data
is the basis to get a meaningful statistical results.
The $\tau$ statistic test is a unique non-parametric technique
widely applied to truncated data for assessing the independence
of parameters \citep{1992ApJ...399..345E}.
The joint operation of the $\tau$ statistic test and
the Lynden-Bell's $C^{-}$ method provides an ideal
nonparametric method to derive the energy function
and event rate of FRB populations.

In this section, we first introduce
the $\tau$ statistic test to remove the dependence between $E$ and $z$.
Then, we describe how to apply the Lynden-Bell's $C^{-}$ method to
independent parameters to obtain the FRB energy function,
redshift distribution and event rate.
The nonparametric analysis notebook and input data for the
non-repeating FRB sample are available on GitHub
\footnote{GitHub:
\url{https://github.com/xxxfei710/Nonpara-FRB-CHIME-Cat2/releases/tag/v1.0.1};
Zenodo:
\dataset[doi:10.5281/zenodo.23531211]
{https://doi.org/10.5281/zenodo.23531211}.}
and archived on Zenodo
\citep{Dong2026Nonpara}.

\subsection{The $\tau$ statistic test}

Although the $C^{-}$ method can overcome data truncation problem,
possible energy-redshift evolution still needs to be
parameterized. Following \cite{2025A&A...698A..18Z}, we adopt the
form of $E \propto (1+z)^{k} $ as an empirical correction, where
$k$ is a constant. Other functional forms could also be adopted in
principle \citep{2025ApJ...988L..64C, 2026ApJ..1003..179J}. Once
the power-law index $k$ is determined, we can correct the energy
as $E' = E/(1+z)^{k}$. Thus, $E'$ and $z$ are independent,
allowing the distribution function to be written as $\Psi(E',z) =
\psi(E') \phi(z)$.

The value of $k$ could be determined from the observational data.
For a specific $k$, the observed data point of each FRB
will change from $(z_i, E_i)$ to the corrected point of $(z_i , E'_i)$.
For the $i$th data point $(z_i, E'_i)$ in the FRB sample,
we first define a data set $J_i$ as
\begin{equation}\label{Ji}
J_i = \{j|E'_j \geq E'_i, z_j\leq z_{i}^{\rm max}\},
\end{equation}
where $E'_i$ is the corrected energy of the $i$th FRB
and $z_{i}^{\rm max}$ is the maximum redshift at which an FRB
with energy $E'_i$ can be detected by CHIME.
The number of FRBs contained in this region is denoted as $n_i$,
and the number of FRBs with redshift $z$ less than or equal to $z_i$
in this region is designated as $R_i$. Then, the statistic $\tau$
is expressed as
\begin{equation}
\tau \equiv \frac{\sum_{i}(R_i - T_i)}{\sqrt{\sum_{i}{V_i}}},
\end{equation}
where $T_i = \frac{1+n_i}{2}$ and $V_i = \frac{(n_i- 1)^2}{12}$
are the expected mean value and the variance of $R_i$, respectively.

According to the $\tau$ statistic test, if $R_i$ is uniformly
distributed between 1 and $n_i$, then the probability of
$R_i \leq T_i$ and $R_i \geq T_i$ should be nearly equal,
so that we have $\tau=0$. In this case, we could know that
the distribution of energy and redshift are independent of each other.
It means that the assumed $k$ value can correctly remove the bias
introduced by the observational selection effects.
On the other hand, if $\tau$ does not equal zero,
then we need to adjust the value of $k$.
The above calculation process is repeated until $\tau = 0$
is satisfied and the correct $k$ value is determined.
Using the $\tau$ statistic test, we finally
get the best value as $k=6.05^{+0.09}_{-0.08}$ for $S_{\rm NR}$,
$k=7.85^{+1.06}_{-0.93}$ for $S_{\rm first}$ and $k=8.72^{+0.83}_{-0.72}$ for $S_{\rm max}$.
The FRB energy is then corrected by
dividing them with $(1+z)^{k}$.

\subsection{The Lynden-Bell's $C^{-}$ method }

The Lynden-Bell's $C^{-}$ method \citep{1971MNRAS.155...95L}
is an effective way
to derive the bivariate distributions of astronomical
objects from the truncated data. In Equation~(\ref{Ji}),
the number of FRBs in the region of $J_i$ is $n_i$.
The Lynden-Bell's $C^{-}$ method does not include
the $i$th FRB in the analysis, which means the FRB number
is smaller by 1, i.e. $N_i = n_i - 1$.

To carry out the calculations, we need to further
define another set as
\begin{equation}\label{JI}
J^{\prime}_i = \{j|E'_{j} \geq E_{i}^{'\rm min}\ , z_j < z_i\},
\end{equation}
where $E_{i}^{'\rm min}$ is the limit energy
at the redshift $z_i$. We denote the number of FRBs
in $J^{\prime}_i$ as $M_i$.
According to the Lynden-Bell's $C^{-}$ method,
the cumulative energy function can then be calculated as
\begin{equation}\label{energyFunction}
\psi(E'_{i}) = \prod\limits_{j>i}(1+\frac{1}{N_j}),
\end{equation}
where $j>i$ means that the operation applies to all FRBs
whose de-evolved energy $E'_{j}$ is larger than $E'_{i}$.
Similarly, the cumulative redshift distribution $\phi(z)$
can be expressed as
\begin {equation}\label {redshiftFunction}
\phi(z_i) = \prod\limits_{j<i}(1+\frac{1}{M_j}),
\end{equation}
where $j<i$ means that the operation applies
to all FRBs whose redshift $z_j$ is less than $z_i$.

The FRB event-rate density can be expressed as:
\begin{equation}\label{formationrate}
\rho(z) = (1+z)\frac{d\phi(z)}{dz}
\left(\frac{dV(z)}{dz}\right)^{-1},
\end{equation}
where the term $(1+z)$ results from the
cosmological time dilation and $dV(z)/dz$ is the
differential comoving volume which can be further expressed as \citep{2019JHEAp..24....1K}
\begin{equation}\label{comovingvolume}
\frac{dV(z)}{dz}=\frac{c}{H_0}\frac{4\pi{D^2_{\rm L}(z)}}{(1+z)^2}\frac{1}{\sqrt{{1 - \Omega_{\rm m}}+\Omega_{\rm m}(1+z)^3}}.
\end{equation}
Note that the comoving volume at a redshift of $z$
is $V = 4 \pi {D_{\rm M}^3} / 3$,
where the comoving distance is $D_{\rm M} = D_{\rm L} / (1+z)$.

\section{results}
\label{sec: RESULTS}

\subsection{Energy function}

\begin{figure*}[htbp]
     \centering
     \includegraphics[width=0.95\textwidth]{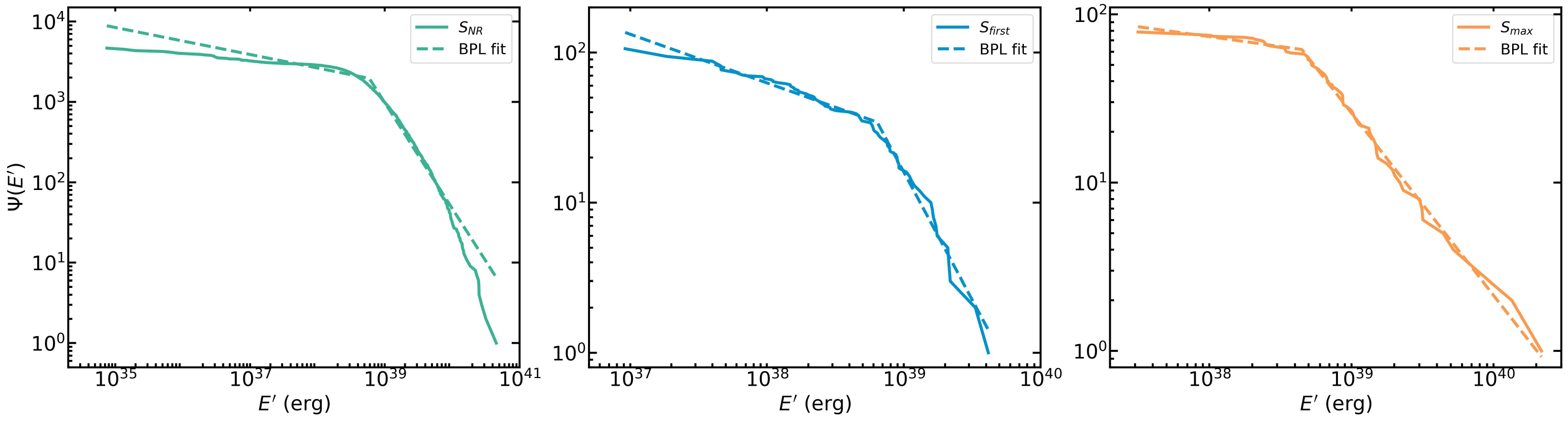}
     \caption{Energy functions after removing
     the redshift evolution.
     The solid line in each panel represents $S_{\rm NR}$,
     $S_{\rm first}$ and $S_{\rm max}$ respectively.
     The dashed line in each panel represents the best fitting result
     by using a broken power-law function.
     }
     \label{fig2}
\end{figure*}

The energy functions of FRBs can be calculated by using
Equation~(\ref{energyFunction}). The results derived from the
three FRB samples are shown in Figure~\ref{fig2}. We see that for
all samples, the energy function decreases with the increasing
de-evolved energy. However, the energy function of the
non-repeating sample is somewhat different from that of the two
repeating samples. A broken power-law function $f(E)$ is applied
to fit the energy functions \citep{2022Natur.609..685X}, i.e.,
\begin{equation}
\label{broken power law}
f(E) \propto \left\{
\begin{array}{ll}
        (E/E_{\rm b})^{\alpha}, \ \
        & E \leq E_{\rm b}, \\
        (E/E_{\rm b})^{\beta}, \ \
        & E > E_{\rm b}
\end{array}
\right.,
\end{equation}
where $E_{\rm b}$ is the energy at the broken position,
$\alpha$ and $\beta$ are the two power-law indices
characterizing the steepness of the energy function
before and after the broken energy.

The fitting results are summarized in Table~\ref{tab:1},
and the best-fit curves are shown as dashed lines in Figure~\ref{fig2}.
The $S_{\rm first}$ sample gives $\chi^2_\nu \approx 1.21$,
consistent with the adopted 10\% uncertainty,
whereas the lower value for $S_{\rm max}$ ($\chi^2_\nu=0.43$)
suggests that its scatter is smaller than assumed.
$S_{\rm first}$ shows the steepest decline toward high energies,
while $S_{\rm max}$ has the flattest distribution and therefore a larger
relative contribution from high-energy bursts.
In contrast, the break energies are broadly similar, indicating a
comparable characteristic energy scale across the samples.

\begin{table}[ht]
    \centering
    \renewcommand{\arraystretch}{1.5}
    \caption{Best fitting parameters of the energy functions.}
    \label{tab:1}
    \begin{tabular}{ccccc}
        \hline
        \hline
        $\mathrm{Sample}$ &
        $\alpha$ &
        $\beta$ &
        $E'_{\rm b}$ (erg) &
        $\mathrm{Reduced}\ \chi^2{}^{*}$ \\
        \hline
        $S_{\rm NR}$ &
        $-0.168 \pm 0.003$ &
        $-1.314 \pm 0.003$ &
        $(5.88 \pm 0.03) \times 10^{38}$ &
        1.88 \\

        $S_{\rm first}$ &
        $-0.33 \pm 0.02$ &
        $-1.70 \pm 0.04$ &
        $(6.38 \pm 0.19) \times 10^{38}$ &
        1.21 \\

        $S_{\rm max}$ &
        $-0.12 \pm 0.04$ &
        $-1.08 \pm 0.02$ &
        $(4.48 \pm 0.20) \times 10^{38}$ &
        0.43 \\
        \hline
    \end{tabular}

    \vspace{2mm}
    \footnotesize
    *The energy function error is taken as 10\%
    when calculating the reduced chi-square.
\end{table}

\subsection{Event rate density}

\begin{figure*}[htbp]
     \centering
     \includegraphics[width=0.5\textwidth]{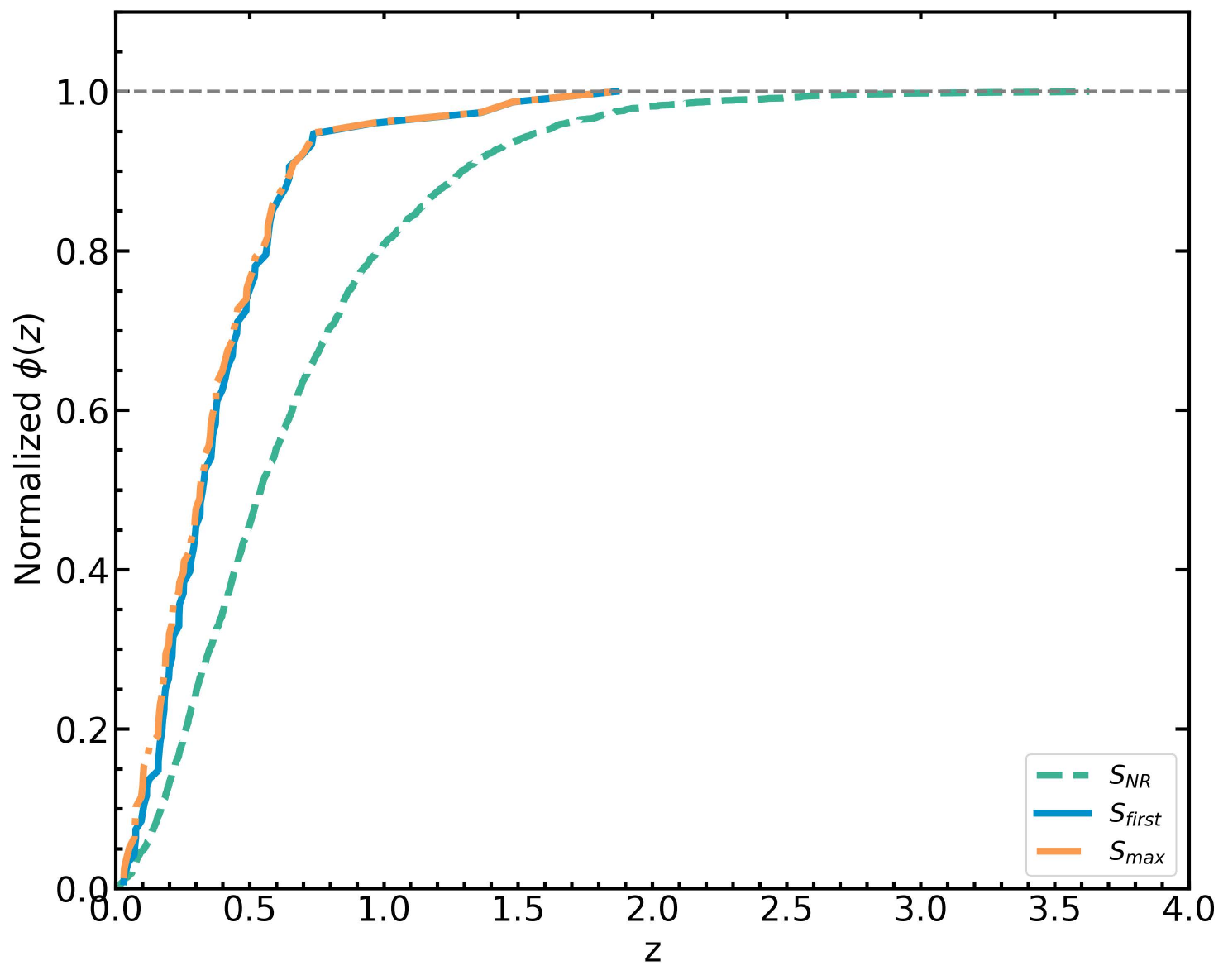}
     \caption{Normalized cumulative redshift distributions.
     The dashed line, solid line, and dashed-dotted lines represent the non-repeating bursts ($S_{\rm NR}$), the first-detected ($S_{\rm first}$) and the highest-fluence ($S_{\rm max}$) bursts of repeating FRBs, respectively.}
     \label{fig3}
\end{figure*}

The normalized cumulative redshift distributions of FRBs
are presented in Figure~\ref{fig3}. The two repeater
samples exhibit consistent distributions,
whereas a clear discrepancy is observed between repeating
and non-repeating populations.
Specifically, the non-repeating FRBs are systematically shifted toward
higher redshifts relative to the repeating samples,
and correspondingly occupy a higher characteristic luminosity range.
The minor offset between the two repeating samples can be
attributed to differences in the adopted redshift and
energy estimates.

\begin{figure}[htbp]
     \centering
     \includegraphics[width=0.5\textwidth]{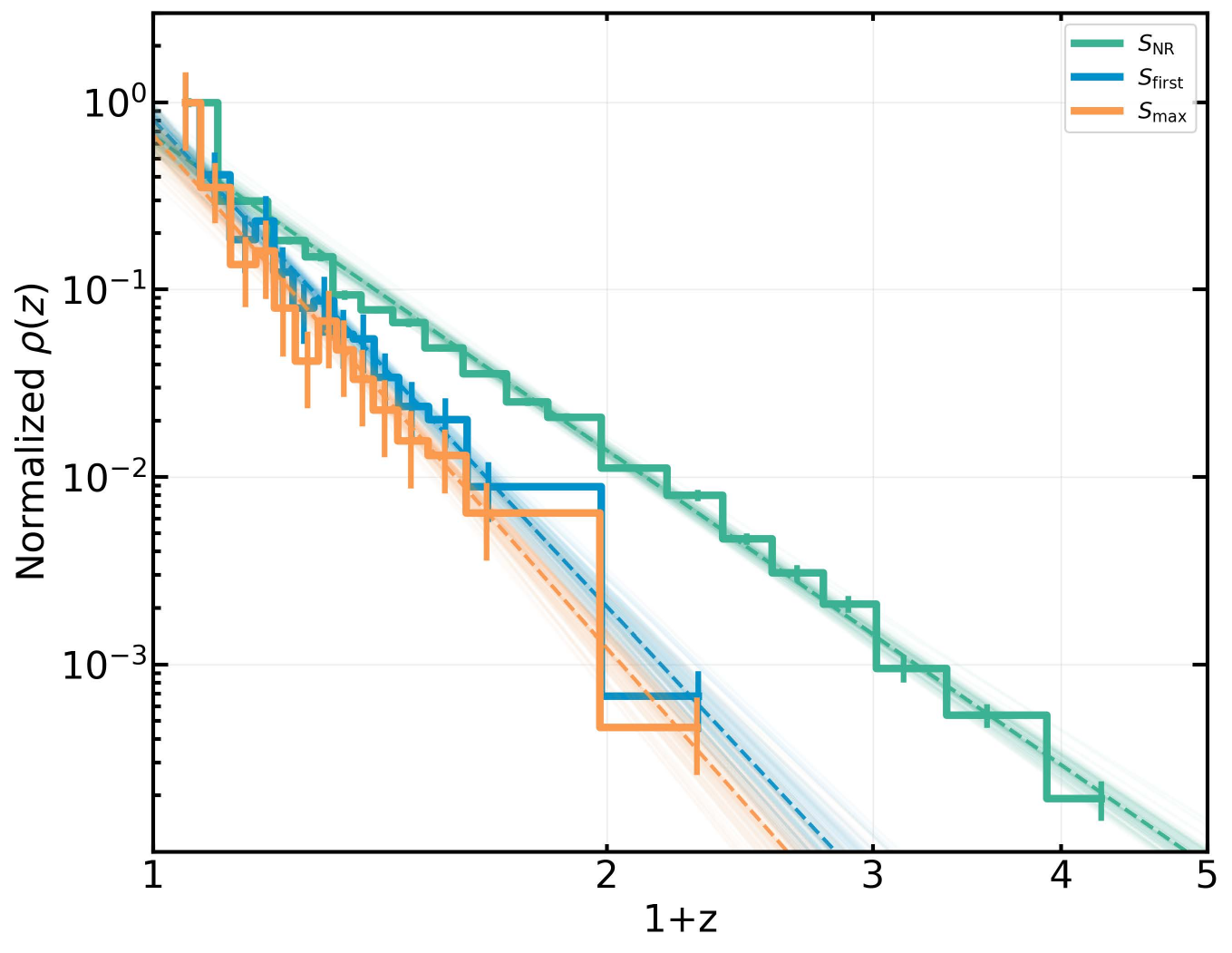}
     \caption{Relative event-rate density $\rho(z)$ as a function of $1+z$. The green, blue, and orange step lines represent the non-repeating sample ($S_{\rm NR}$), the first-detected bursts ($S_{\rm first}$), and the highest-fluence bursts ($S_{\rm max}$) of repeating FRBs, respectively. The event-rate density of each sample is independently normalized to its maximum value. The dashed lines show the best-fit power-law models, with shaded regions indicating the corresponding $1\sigma$ confidence intervals.}
     \label{fig4}
\end{figure}

The derived event rates $\rho(z)$ are shown in Figure~\ref{fig4}.
All samples show a monotonic decline with increasing redshift,
spanning nearly four orders of magnitude.
This evolution is well described by a power-law function,
$\rho(z) \propto (1+z)^B$. The best-fit results
by the Markov Chain Monte Carlo analysis are overplotted
in Figure~\ref{fig4}.
For the $S_{\rm NR}$ sample, we obtain $B = -5.57^{+0.15}_{-0.15}$,
consistent with previous results of non-repeating bursts based on the
CHIME Catalog 1 ($-4.9\pm0.3$; \citealt{2024ApJ...973L..54C})
and recent work of Catalog 2 ($-5.38\pm0.02$; \citealt{2026ApJ..1003..179J}).

For the repeating samples, the best-fit indices are $B =
-8.63^{+0.46}_{-0.41}$ for $S_{\rm first}$ and $B =
-9.10^{+0.55}_{-0.56}$ for $S_{\rm max}$. The two repeating samples exhibit consistent evolutionary trends, suggesting similar population properties. In contrast, their inferred evolution is significantly steeper than that of the non-repeating sample. This difference may point to distinct underlying populations.

\subsection{Comparison to SFR and SMD}

\begin{figure}[htbp]
     \centering
     \includegraphics[width=0.5\textwidth]{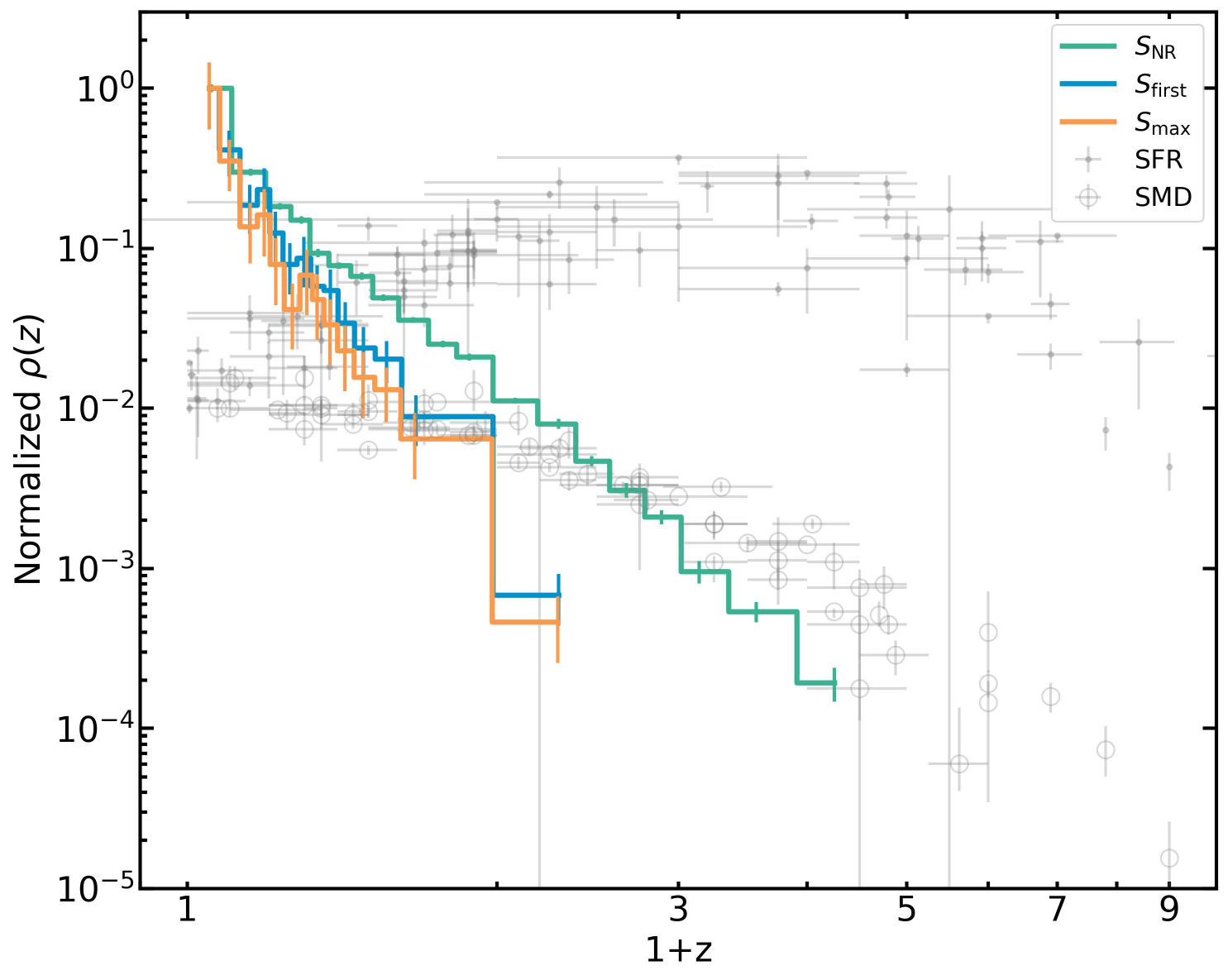}
     \caption{Comparison of the cosmic SFR, SMD, and FRB event rates. The step lines are identical to those shown in Figure~\ref{fig4}. The filled circles denote the SFR measurements \citep{2006ApJ...651..142H, 2007ApJ...657..669T, 2011Natur.469..504B, 2008MNRAS.388.1487L}, while the hollow circles represent the SMD data \citep{2014ARA&A..52..415M}. The SFR and SMD are scaled by arbitrary factors for comparison.}
     \label{fig5}
\end{figure}

Different FRB progenitor channels may exhibit distinct
cosmic event-rate evolution, depending on their delay times
relative to star formation \citep{2023RvMP...95c5005Z}. Young
progenitors are expected to approximately trace the SFR
\citep{2021ApJ...907L..31B, 2022MNRAS.510L..18J,
2026arXiv260709109W}, whereas older populations may evolve more
closely with the stellar mass density (SMD;
\citealt{2022MNRAS.511.1961H, 2024ApJ...973L..54C}) or reflect a
range of delay times \citep{2022ApJ...924L..14Z,
2026arXiv260704792Z}. Comparing FRB event-rate evolution with the
SFR and SMD may therefore help constrain the underlying
populations.

As seen in Figure~\ref{fig5}, none of the FRB samples closely
trace the SFR. Although the FRB event rates decline with redshift,
qualitatively resembling the SMD, significant deviations remain.
Among the samples, only the non-repeating population shows
moderate agreement with the SMD at $z \gtrsim 0.8$.
These results are consistent with recent analyses of the
CHIME Catalog 1 \citep{2024ApJ...973L..54C, 2025ApJ...988L..64C,
2025arXiv250912016P} and Catalog 2 \citep{2026ApJ..1003..179J},
which suggest that FRBs are preferentially associated with
older stellar populations, such as neutron stars or black holes.

In particular, \cite{2025arXiv250912016P} divided repeating FRBs
into high-/low-frequency and high-/low-luminosity subsamples and
found that all subsamples show event rates exceeding the SFR at
$z<1$, with the rates generally decreasing toward higher redshift
in this range. This behavior is especially evident for
low-frequency FRBs. Our repeating samples derived from CHIME
(400--800 MHz) are consistent with their results, suggesting that
repeating FRBs do not simply trace the cosmic star formation
history. \cite{2025ApJ...988L..64C} further showed that
uncertainties in DM-based redshift estimates do not significantly
affect this low-redshift excess. Additional observational
evidence, such as the lack of a persistent radio counterpart for
FRB 20250316A \citep{2025ApJ...991L..20A}, also favors an origin
in older, low-density environments.

At the same time, FRBs do not appear to be unbiased tracers
of star formation. \cite{2024Natur.635...61S}
showed that FRBs preferentially occur in massive star-forming
galaxies, unlike core-collapse supernovae that closely
follow the SFR. A mixed-population scenario has therefore
been proposed, in which some FRBs trace recent star formation
while others track stellar mass \citep{2025ApJS..277...43M, 2025ApJ...991...85L, 2026ApJ...996...78H}.
Our results are also consistent with such a heterogeneous population scenario, in which repeating and apparently non-repeating FRBs may represent distinct or partially distinct subpopulations.

\section{Robustness Tests}
\label{sec:test}
\subsection{Sensitivity to the Fluence Threshold}
We approximate the CHIME selection function using a hard fluence threshold for each sample, although the actual detection efficiency also depends on declination, burst width, and other factors \citep{2021ApJS..257...59C, 2026arXiv260626334M, 2026arXiv260704792Z}. Since the truncation boundary determines the associated sets in the Lynden-Bell's $C^{-}$ method \citep{2021MNRAS.504.4192B}, we examine the sensitivity of our results to the adopted thresholds.

We systematically varied the fluence threshold $F_{\rm lim}$ for each sample, ensuring that at least 90\% of the original events were retained at every threshold. Specifically, we adopted 100 thresholds randomly over the ranges 0.1--0.9, 0.1--1.2, and 0.1--3.8 Jy ms for $S_{\rm NR}$, $S_{\rm first}$, and $S_{\rm max}$, respectively. For each threshold, we repeated the full analysis to get the energy-evolution parameter $k$, the intrinsic energy function and the rate-evolution index $B$.

\begin{figure}[htbp]
    \centering
    \includegraphics[width=0.4\textwidth]{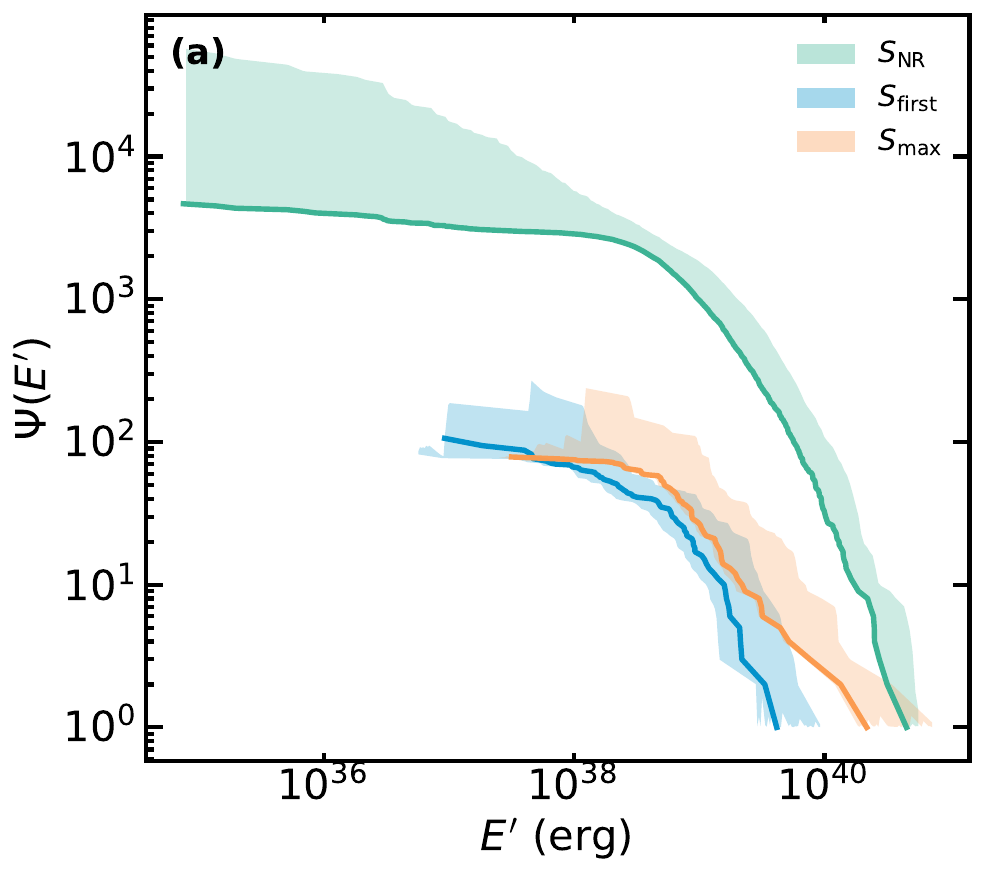}
    \centering
    \includegraphics[width=0.4\textwidth]{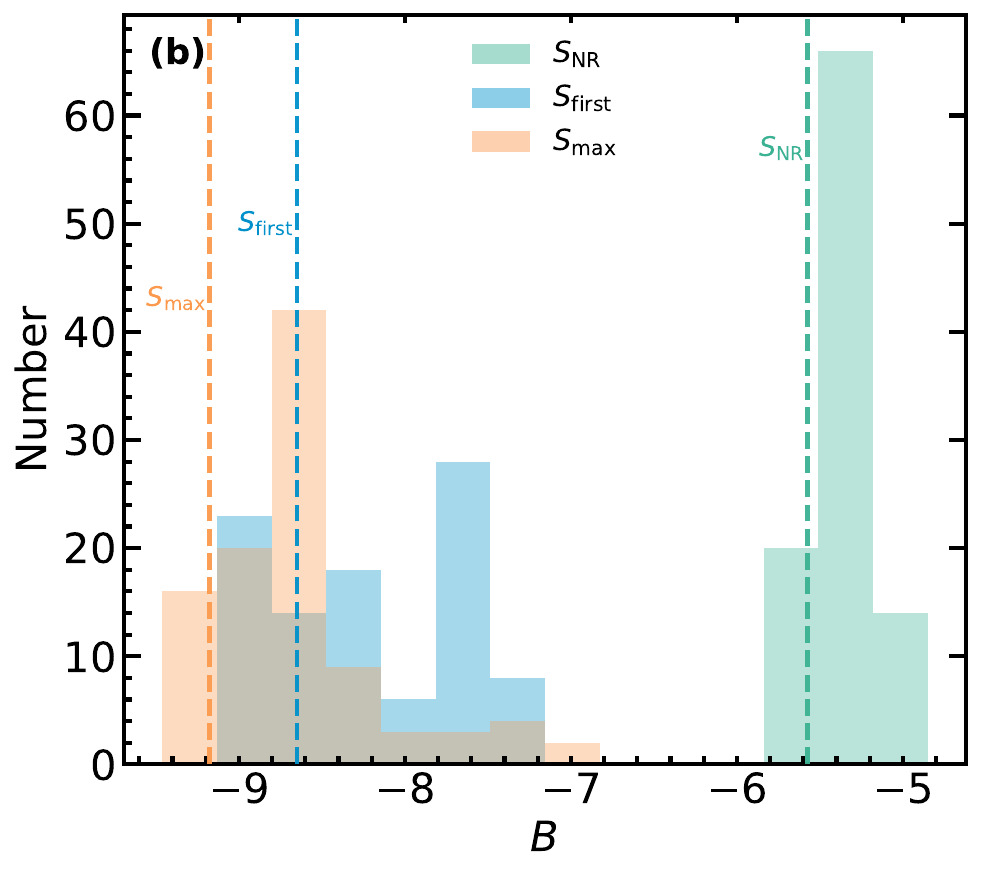}
    \caption{Robustness tests for the adopted fluence thresholds.
    Panel (a): Reconstructed intrinsic energy functions. The solid lines represent the results of the original thresholds, while the shaded regions indicate the full ranges obtained from the 100 threshold configurations.
    Panel (b): Distributions of the fitted rate-evolution index $B$ across the 100 threshold configurations for each sample. The vertical dashed lines indicate the corresponding results from the original analysis.}
    \label{fig6}
\end{figure}

The inferred energy-evolution index $k$ decreases with
increasing fluence threshold, ranging from 6.05 to 4.60, 9.02 to
5.26, and 9.12 to 5.26 for $S_{\rm NR}$, $S_{\rm first}$, and
$S_{\rm max}$, respectively, indicating its sensitivity to the
adopted threshold, consistent with \cite{2021MNRAS.504.4192B}.
Figure~\ref{fig6} summarizes the corresponding variations in the
energy functions and rate-evolution indices. Panel (a) shows that
the reconstructed energy functions vary, particularly at low
energies, while retaining overall differences between
non-repeaters and repeaters. More importantly, panel (b) shows
that the inferred rate-evolution indices $B$ of the two repeating
samples substantially overlap, which remain clearly separated from
the $S_{\rm NR}$ sample.

\subsection{Effect of Sample Size on the Event Rate}

Most FRBs are detected as one-off events, resulting in a
substantially larger sample size for $S_{\rm NR}$ when compared to
$S_{\rm first}$ and $S_{\rm max}$. This disparity raises the
question of whether the inferred event-rate evolution is biased by
sample size. To quantify this effect, we perform stratified random
sampling on $S_{\rm NR}$. We generate 2000 mock samples, each
randomly drawn from $S_{\rm NR}$ and matched in size to the
repeating samples. For each realization, we compute the event rate
using the same nonparametric method and fit it with a power-law
model.

\begin{figure}[htbp]
     \centering
     \includegraphics[width=0.5\textwidth]{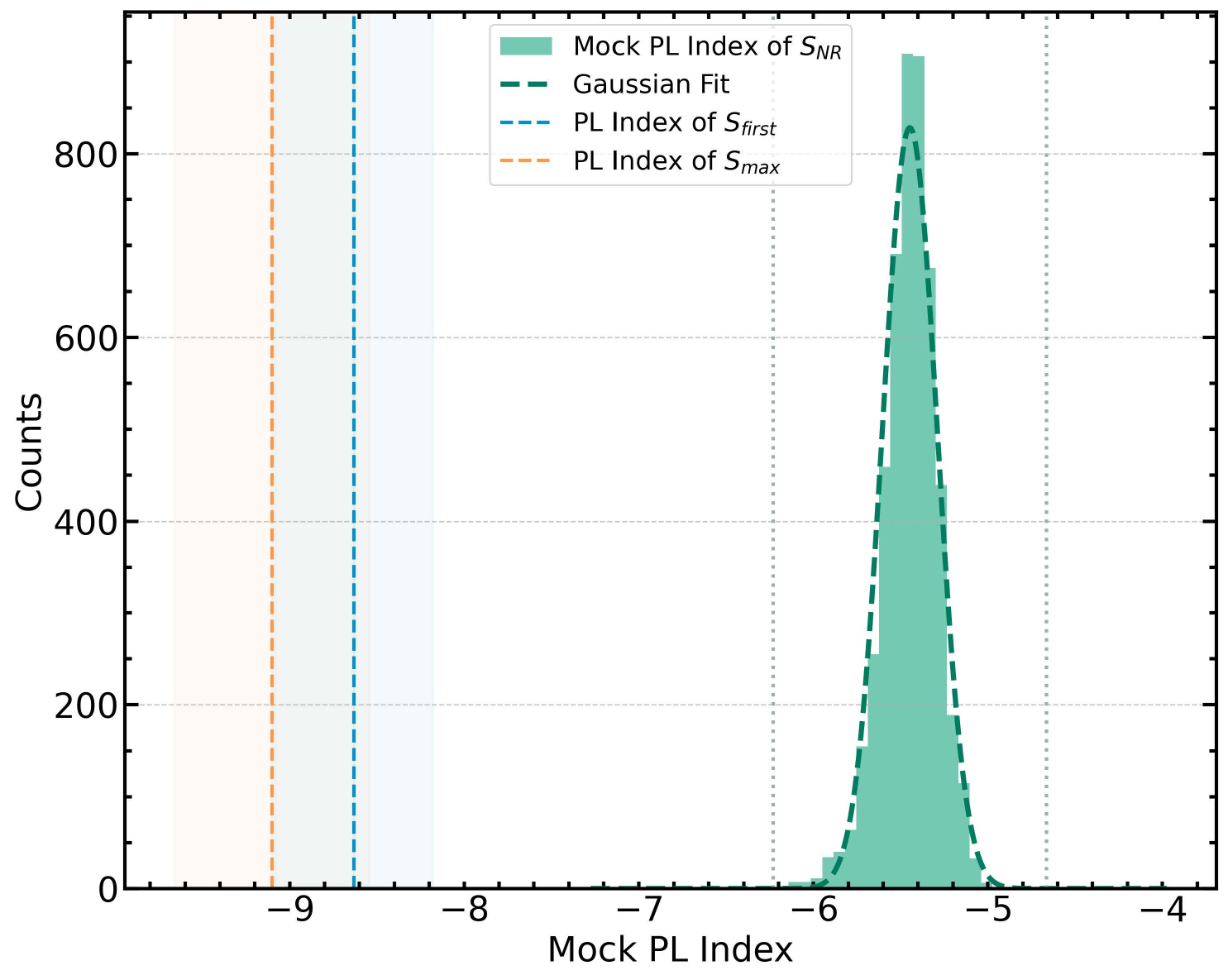}
     \caption{Distribution of the fitted power-law indices
     derived from the stratified random sampling simulations.
     The green dashed line shows the Gaussian fit to the
     distribution. The vertical dashed lines mark the best-fit
     indices of $S_{\rm first}$ and $S_{\rm max}$,
     with the shaded regions indicating their corresponding
     uncertainties. The vertical dotted line denotes the
     $5\sigma$ offset from the Gaussian mean.}
     \label{fig7}
\end{figure}

The distribution of the resulting indices is shown in
Figure~\ref{fig7}. A Gaussian fit yields a mean of $-5.45$ with a
standard deviation of $0.16$, similar to the best-fit value of the
full $S_{\rm NR}$ sample ($-5.57$). In contrast, the indices of
the repeating samples, $-8.63$ ($S_{\rm first}$) and $-9.10$
($S_{\rm max}$), are well outside the $5\sigma$ range of the mock
distribution. It indicates that the inferred difference in
the rate-evolution index $B$ is unlikely to arise solely from
finite-sample fluctuations under the adopted assumptions.

\subsection{Systematic Uncertainty Tests}
\label{DMtest}
\begin{figure}[htbp]
     \centering
     \includegraphics[width=0.98\textwidth]{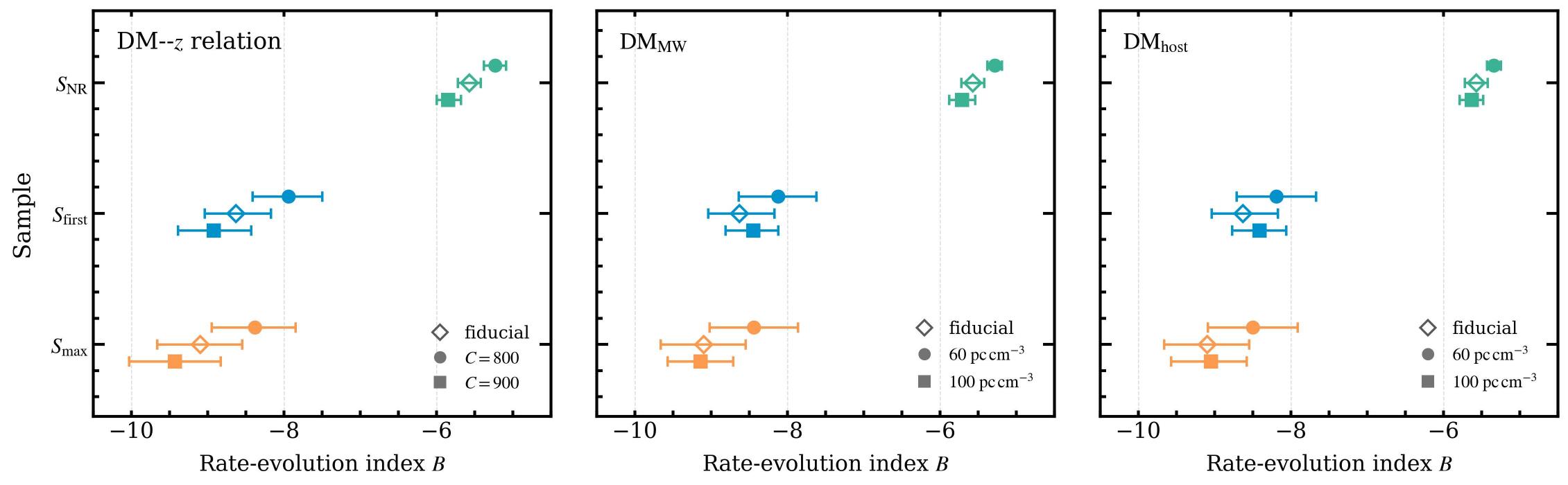}
     \caption{The cosmic rate-evolution index $B$
     under different assumptions for the DM-redshift conversion.
     Different colors represent the three FRB samples. Open
     diamonds indicate the fiducial results, while circles and
     squares denote the cases with one parameter among $C, \rm DM_{\rm MW}$
     and $\rm DM_{\rm host}$ altered in each panel.}
     \label{fig8}
\end{figure}

In our analysis, the inferred FRB redshifts depend on the
adopted DM--$z$ relation (featured by $C=855$; Equation~\ref{z}),
as well as the $\rm DM_{\rm MW}$ and $\rm DM_{\rm host}$ parameters, which
are adopted as $\rm DM_{\rm MW}=80\,pc\,cm^{-3}$ and $\rm DM_{\rm
host}=84\,pc\,cm^{-3}$ in our modeling. We therefore need to
examine how uncertainties in the DM--$z$ relation and relevant
parameters affect our results. To do so, we have tried other
parameter values, i.e. $C=800$ or $900$, $\rm DM_{\rm MW}=60$ or
$100\,\rm pc\,cm^{-3}$, and $\rm DM_{\rm host}=60$ or $100\,\rm
pc\,cm^{-3}$. We vary one parameter at a time while keeping the
other parameters at their fiducial values.

The inferred energy-evolution parameter $k$ varies with
the adopted parameters, ranging from 5.78 to 6.26 for $S_{\rm
NR}$, 6.79 to 7.73 for $S_{\rm first}$, and 7.80 to 8.98 for
$S_{\rm max}$. It indicates that the inferred energy evolution is
affected by the adopted assumptions. The resulting cosmic
rate-evolution indices $B$ are shown in Figure~\ref{fig8}.
Although $B$ also varies with the adopted parameters, the
differences between $S_{\rm NR}$ and the two repeating samples
persist. Meanwhile, $S_{\rm first}$ and $S_{\rm max}$ remain
broadly consistent within their uncertainties. It clearly shows
that repeating FRBs and apparently non-repeating FRBs are
different at population level.

\section{Conclusions and Discussion}
\label{sec: Conclusion and Discussion}

Using the newly released CHIME/FRB Catalog 2, we compare
apparently non-repeating FRBs with the first-detected and highest-
fluence bursts from repeating sources using a non-parametric
method. The two repeating samples exhibit consistent
energy functions and cosmic event-rate evolution, while both show
different results from the non-repeating sample. The differences
in the inferred event-rate evolution are robust under various
tests, including matched-size resampling, varying fluence
threshold and different DM--redshift conversion assumptions.
Furthermore, the inferred event-rate evolution of all the three
samples deviates from the cosmic star formation history. The
results suggest that repeating and apparently non-repeating FRBs
should be different populations.

Sample selection and completeness corrections are crucial
for inferring FRB event-rate evolution. Using the Lynden-Bell's
$C^{-}$ method, \citet{2024ApJ...973L..54C} found declining rates
across different fluence thresholds. \cite{2025ApJ...988L..64C}
and \cite{2026ApJ..1003..179J} adopted a smoothly broken power-law
energy-redshift relation and reported similar declines under
different DM-based redshift estimates. In contrast,
\citet{2025A&A...698A..18Z} constructed three volume-limited
subsamples with different evolutionary behaviors, including little
or no evolution in two of them. The $V_{\max}$ analysis of
\citet{2022MNRAS.511.1961H} also found declining rates, whereas
forward modeling with more detailed instrumental selection effects
could not rule out SFR-tracing evolution
\citep{2025A&A...698A.127M}. Using the same non-parametric $C^{-}$
method, we recover the declining rate evolution of non-repeaters.
More importantly, the first-detected and highest-fluence bursts of
repeaters exhibit distinct energy functions and rate evolution
when compared to non-repeaters.

Whether repeating and apparently non-repeating FRBs
represent distinct populations is an important problem \citep{2018ApJ...854L..12P, 2018NatAs...2..839C, 2021ApJ...906L...5A}. Previous
studies have found differences in their burst morphology and
luminosity functions \citep{2020MNRAS.494.2886H,
2021ApJ...923....1P}, while follow-up observations suggest that
some non-repeaters may have very low activity
\citep{2025MNRAS.540.3709U, 2025ApJ...989L..48C,
2025ApJ...995....8L, 2026ApJ...996...47Y}. Machine-learning
studies have also identified differences between the two classes
\citep{2023MNRAS.518.1629L, 2023MNRAS.519.1823Z, 2024ApJ...977..273G}, and recent
simulation-based inference favors a distinct one-off subpopulation
\citep{2026arXiv261018710S}. However, there are also other studies
supporting a common population with a broad range of repetition
rates \citep{2024MNRAS.52711158Y, 2025ApJ...993...37B}. It has
been argued that some differences may arise from selection effects
\citep{2026SCPMA..6909512S}. Our results provide complementary
evidence from energy functions and cosmic event-rate evolution,
suggesting possible differences at the population level.

As repeater identification depends on the repetition rate
and source activity, as well as on the observation cadence and
telescope sensitivity, some apparently non-repeating FRBs may
actually be unidentified repeaters, introducing uncertainty into
our sample classification \citep{2023PASA...40...57J,
2023ApJ...947...83C}. Moreover, brighter bursts may be detected
from repeaters in future observations, which would naturally
change our highest-fluence sample and potentially affect the
inferred population differences. It is expected that more
observations in the future will help us overcome these problems
and let us be more close to the reality.

Recent FRB studies have shown that inferred redshift distributions
are affected by selection effects, redshift uncertainties, and
possible luminosity- or energy-function evolution
\citep{2026arXiv260704792Z, 2026arXiv260709109W}. We therefore do
not interpret the derived event-rate evolution as a unique
physical delay-time distribution. Our main result is comparative:
when analyzed with the same non-parametric method and threshold
prescription, repeating and non-repeating FRBs show systematically
different energy functions and event-rate evolution. These
uncertainties may affect the detailed rate shape and
normalization, but the relative comparison is less sensitive to
them.

\section{Acknowledgements}
We thank the anonymous referee for the useful comments and
suggestions that led to the overall improvement of the
presentation. This work is supported by the National Natural
Science Foundation of China (Grant No. 12233002, 12273113,
12573051), by the National Key R\&D Program of China (2025SKA0140100), by the Shandong Provincial Natural Science Foundation
(No. ZR2023MA049), and the Major Science and Technology Program of
Xinjiang Uygur Autonomous Region (No. 2022A03013-1). J.-J.G. acknowledges support from the Youth Innovation Promotion Association (2023331).

\bibliography{refs}

\clearpage
\centering
\startlongtable
\begin{deluxetable}{ccccccccccc}
\tabletypesize{\scriptsize}
\renewcommand{\arraystretch}{0.95}

\tablecaption{Key parameters of the 78 CHIME repeating FRBs.\label{tab:2}}

\tablewidth{0pt}

\tablehead{
\colhead{$\mathrm{FRB}$} &
\colhead{$\mathrm{ID}_{\rm m}$} &
\colhead{${z}_{\rm m}$} &
\colhead{${F}_{\rm m}$} &
\colhead{$\nu_{\mathrm{m}}$} &
\colhead{${E}_{\mathrm{iso,m}}$} &
\colhead{$\mathrm{ID}_{\rm f}$} &
\colhead{${z}_{\rm f}$} &
\colhead{${F}_{\rm f}$} &
\colhead{$\nu_{\mathrm{f}}$} &
\colhead{${E}_{\mathrm{iso,f}}$}
\\
\colhead{} &
\colhead{} &
\colhead{} &
\colhead{[$\mathrm{Jy\ ms}$]} &
\colhead{[$\mathrm{MHz}$]} &
\colhead{[$10^{39}\ \mathrm{erg}$]} &
\colhead{} &
\colhead{} &
\colhead{[$\mathrm{Jy\ ms}$]} &
\colhead{[$\mathrm{MHz}$]} &
\colhead{[$10^{39}\ \mathrm{erg}$]}
\\
\colhead{(1)} &
\colhead{(2)} &
\colhead{(3)} &
\colhead{(4)} &
\colhead{(5)} &
\colhead{(6)} &
\colhead{(7)} &
\colhead{(8)} &
\colhead{(9)} &
\colhead{(10)} &
\colhead{(11)}
}

\startdata
20171019A & 46707700 & 0.38  & 22.23  & 632.24  & 51.26  & 46707700 & 0.38  & 22.23  & 632.24  & 51.26  \\
20180814A & 19815274 & 0.03  & 32.28  & 430.55  & 0.32  & 10889573 & 0.03  & 1.06  & 462.90  & 0.01  \\
20180908B & 12806028 & 0.04  & 6.81  & 591.93  & 0.15  & 12806028 & 0.04  & 6.81  & 591.93  & 0.15  \\
20180910A & 94585180 & 0.66  & 12.12  & 515.81  & 73.04  & 13168374 & 0.65  & 5.60  & 608.24  & 38.01  \\
20180916B & 23491552 & 0.23  & 52.78  & 672.86  & 48.26  & 15372046 & 0.24  & 6.93  & 610.05  & 5.79  \\
20181017A & 30749404 & 1.36  & 49.58  & 491.10  & 1199.52  & 18572783 & 1.36  & 16.00  & 524.67  & 413.15  \\
20181018B & 18594059 & 0.17  & 7.90  & 572.25  & 2.99  & 18594059 & 0.17  & 7.90  & 572.25  & 2.99  \\
20181119A & 174845492 & 0.25  & 9.54  & 632.04  & 9.69  & 20854062 & 0.25  & 4.25  & 495.62  & 3.36  \\
20181128A & 21538821 & 0.36  & 8.10  & 474.91  & 12.84  & 21538821 & 0.36  & 8.10  & 474.91  & 12.84  \\
20181201D & 148893983 & 0.35  & 7.73  & 646.47  & 15.84  & 21841625 & 0.36  & 1.15  & 400.21  & 1.51  \\
20181226F & 29110534 & 0.10  & 2.10  & 475.06  & 0.23  & 29110534 & 0.10  & 2.10  & 475.06  & 0.23  \\
20190110C & 60955229 & 0.07  & 2.56  & 440.17  & 0.15  & 25677682 & 0.07  & 1.92  & 422.79  & 0.10  \\
20190113A & 77799213 & 0.33  & 6.90  & 536.32  & 10.30  & 26354477 & 0.33  & 6.59  & 728.94  & 13.54  \\
20190116A & 26804245 & 0.36  & 4.11  & 649.35  & 8.57  & 26804245 & 0.36  & 4.11  & 649.35  & 8.57  \\
20190117A & 197381138 & 0.29  & 47.95  & 474.23  & 49.16  & 27052203 & 0.29  & 11.23  & 503.89  & 11.93  \\
20190208A & 77049697 & 0.52  & 13.54  & 439.20  & 42.26  & 29889091 & 0.52  & 3.84  & 465.73  & 12.71  \\
20190208C & 170809686 & 0.10  & 25.21  & 422.13  & 2.27  & 29866131 & 0.10  & 1.87  & 475.92  & 0.19  \\
20190209A & 30112019 & 0.33  & 4.90  & 464.78  & 6.35  & 30015540 & 0.33  & 3.37  & 470.42  & 4.35  \\
20190212A & 65300794 & 0.18  & 6.20  & 523.21  & 2.42  & 30292842 & 0.18  & 2.93  & 471.33  & 1.03  \\
20190213A & 38776825 & 0.61  & 4.10  & 538.07  & 21.60  & 30351693 & 0.61  & 2.26  & 434.80  & 9.62  \\
20190222A & 31468495 & 0.37  & 13.34  & 444.86  & 21.16  & 31468495 & 0.37  & 13.34  & 444.86  & 21.16  \\
20190224D & 68252213 & 0.72  & 7.81  & 550.16  & 59.84  & 31644043 & 0.73  & 6.38  & 530.82  & 48.40  \\
20190226B & 69262405 & 0.58  & 3.67  & 562.93  & 18.68  & 31851655 & 0.58  & 2.22  & 586.19  & 11.77  \\
20190303A & 61303573 & 0.07  & 13.36  & 527.19  & 0.89  & 32329698 & 0.07  & 2.28  & 635.02  & 0.19  \\
20190320E & 33928716 & 0.17  & 12.72  & 628.65  & 5.74  & 33928716 & 0.17  & 12.72  & 628.65  & 5.74  \\
20190417A & 157365378 & 1.48  & 17.65  & 480.11  & 487.14  & 37085135 & 1.48  & 3.68  & 561.15  & 118.66  \\
20190430C & 38440671 & 0.30  & 6.85  & 660.75  & 10.17  & 38440671 & 0.30  & 6.85  & 660.75  & 10.17  \\
20190604A & 40227427 & 0.49  & 8.84  & 428.77  & 23.51  & 40227427 & 0.49  & 8.84  & 428.77  & 23.51  \\
20190609C & 140210071 & 0.40  & 3.62  & 437.51  & 6.43  & 41252817 & 0.40  & 1.20  & 420.23  & 2.05  \\
20190716C & 206185996 & 0.21  & 3.81  & 695.98  & 2.86  & 206185996 & 0.21  & 3.81  & 695.98  & 2.86  \\
20190804E & 46439431 & 0.25  & 6.46  & 436.17  & 4.45  & 46439431 & 0.25  & 6.46  & 436.17  & 4.45  \\
20190905A & 183226337 & 0.10  & 20.80  & 404.16  & 2.05  & 52473268 & 0.12  & 1.81  & 497.62  & 0.28  \\
20190907A & 105314031 & 0.19  & 6.95  & 511.29  & 2.96  & 52839758 & 0.19  & 0.48  & 622.30  & 0.25  \\
20190915D & 72987857 & 0.41  & 13.87  & 490.95  & 29.16  & 54720885 & 0.41  & 13.39  & 536.00  & 30.84  \\
20190929E & 195657636 & 0.44  & 18.49  & 458.92  & 42.64  & 56556394 & 0.44  & 3.96  & 433.51  & 8.52  \\
20191013D & 84271986 & 0.45  & 22.19  & 443.05  & 51.65  & 57583131 & 0.45  & 11.34  & 678.31  & 40.73  \\
20191106C & 145408666 & 0.22  & 4.29  & 420.69  & 2.05  & 60792403 & 0.21  & 1.68  & 644.54  & 1.21  \\
20191113C & 75647643 & 0.57  & 2.39  & 741.52  & 15.07  & 61706633 & 0.57  & 1.26  & 684.88  & 7.28  \\
20191114A & 81130446 & 0.49  & 6.71  & 528.41  & 22.14  & 61913156 & 0.49  & 4.69  & 470.01  & 13.62  \\
20200118D & 95677738 & 0.58  & 3.02  & 480.41  & 12.73  & 69246471 & 0.58  & 0.95  & 408.80  & 3.40  \\
20200127B & 70836867 & 0.24  & 5.17  & 438.22  & 3.17  & 70836867 & 0.24  & 5.17  & 438.22  & 3.17  \\
20200202A & 138244000 & 0.70  & 25.77  & 585.83  & 196.78  & 71571712 & 0.69  & 8.49  & 527.60  & 57.58  \\
20200223B & 95822946 & 0.05  & 9.55  & 692.93  & 0.39  & 74317193 & 0.05  & 1.49  & 719.64  & 0.06  \\
20200308B & 96534808 & 0.07  & 8.86  & 418.20  & 0.41  & 76199216 & 0.07  & 5.18  & 529.02  & 0.30  \\
20200420A & 118315775 & 0.63  & 22.38  & 697.67  & 164.92  & 81358007 & 0.63  & 1.67  & 444.03  & 7.91  \\
20200510A & 83744922 & 0.16  & 22.11  & 400.21  & 5.58  & 83744922 & 0.16  & 22.11  & 400.21  & 5.58  \\
20200619A & 218607166 & 0.35  & 3.34  & 449.45  & 4.62  & 94243065 & 0.35  & 0.41  & 442.42  & 0.56  \\
20200809E & 156747973 & 1.86  & 6.60  & 526.99  & 306.42  & 114185370 & 1.86  & 2.01  & 402.01  & 71.20  \\
20200905B & 132687967 & 0.18  & 7.64  & 485.17  & 2.92  & 132687967 & 0.18  & 7.64  & 485.17  & 2.92  \\
20200913C & 134712756 & 0.51  & 2.10  & 520.28  & 7.59  & 134432536 & 0.52  & 1.32  & 482.75  & 4.47  \\
20200926A & 307983650 & 0.74  & 15.13  & 486.39  & 107.15  & 136489401 & 0.74  & 3.17  & 429.19  & 19.80  \\
20200929C & 136785433 & 0.32  & 9.49  & 431.34  & 10.28  & 136785433 & 0.32  & 9.49  & 431.34  & 10.28  \\
20201114A & 143460991 & 0.20  & 4.13  & 622.11  & 2.50  & 143460991 & 0.20  & 4.13  & 622.11  & 2.50  \\
20201124A & 172176403 & 0.31  & 129.88  & 451.77  & 144.83  & 144650722 & 0.31  & 6.42  & 465.56  & 7.48  \\
20201130A & 166028328 & 0.16  & 11.59  & 513.78  & 3.55  & 145286975 & 0.16  & 7.30  & 642.88  & 2.84  \\
20201221B & 163213672 & 0.44  & 9.05  & 428.04  & 19.06  & 148353736 & 0.43  & 1.22  & 449.16  & 2.68  \\
20210125A & 213890320 & 0.37  & 6.47  & 444.42  & 10.26  & 153736954 & 0.37  & 1.81  & 465.17  & 2.98  \\
20210209C & 172091258 & 0.28  & 3.98  & 741.71  & 5.65  & 157073938 & 0.28  & 2.85  & 515.32  & 2.81  \\
20210219C & 168239699 & 0.45  & 243.54  & 437.68  & 572.35  & 160520504 & 0.45  & 1.17  & 445.03  & 2.78  \\
20210305A & 299682819 & 0.12  & 3.19  & 527.04  & 0.56  & 163704771 & 0.12  & 0.71  & 486.95  & 0.11  \\
20210323C & 166470978 & 0.16  & 6.92  & 489.71  & 2.06  & 165567608 & 0.16  & 1.99  & 512.15  & 0.62  \\
20210330B & 200258236 & 0.30  & 6.16  & 520.91  & 7.04  & 200258236 & 0.30  & 6.16  & 520.91  & 7.04  \\
20210409B & 285831667 & 0.65  & 14.41  & 518.03  & 83.84  & 167450017 & 0.65  & 4.48  & 723.47  & 36.21  \\
20210504D & 170103438 & 0.50  & 13.39  & 587.15  & 51.74  & 170103438 & 0.50  & 13.39  & 587.15  & 51.74  \\
20210601A & 172714732 & 0.32  & 14.33  & 668.30  & 24.93  & 172714723 & 0.32  & 8.86  & 608.00  & 14.20  \\
20210822E & 209838881 & 0.20  & 40.35  & 400.21  & 15.41  & 181654027 & 0.20  & 1.61  & 427.09  & 0.65  \\
20210924C & 191642372 & 0.42  & 4.05  & 569.98  & 10.39  & 187695669 & 0.42  & 0.98  & 529.24  & 2.33  \\
20211030A & 194297475 & 0.21  & 5.82  & 412.76  & 2.63  & 194297475 & 0.21  & 5.82  & 412.76  & 2.63  \\
20211120C & 226660981 & 0.57  & 7.72  & 400.21  & 26.30  & 198377378 & 0.57  & 1.81  & 646.45  & 10.05  \\
20211227A & 204660530 & 0.97  & 6.13  & 590.25  & 91.08  & 204660530 & 0.97  & 6.13  & 590.25  & 91.08  \\
20220101A & 234763312 & 0.28  & 1.80  & 648.33  & 2.30  & 205209223 & 0.28  & 1.65  & 480.33  & 1.56  \\
20220207A & 210456524 & 0.13  & 54.00  & 417.03  & 8.47  & 210456524 & 0.13  & 54.00  & 417.03  & 8.47  \\
20220316A & 216115068 & 0.24  & 0.62  & 443.40  & 0.38  & 216115068 & 0.24  & 0.62  & 443.40  & 0.38  \\
20220319C & 230162661 & 0.19  & 17.00  & 427.58  & 6.12  & 216787925 & 0.18  & 2.59  & 432.36  & 0.89  \\
20220505A & 289479306 & 0.55  & 22.17  & 400.21  & 71.26  & 224160869 & 0.56  & 3.67  & 730.41  & 22.17  \\
20220529A & 280718945 & 0.10  & 16.52  & 489.73  & 2.08  & 227531533 & 0.11  & 3.77  & 448.35  & 0.44  \\
20220618C & 243462559 & 0.03  & 70.30  & 419.40  & 0.61  & 231150454 & 0.03  & 27.27  & 514.98  & 0.28  \\
20220912A & 252450889 & 0.07  & 123.84  & 552.97  & 8.09  & 244765186 & 0.07  & 1.47  & 456.24  & 0.08  \\
\enddata

\tablecomments{
(1): the names of repeating FRBs.
(2)--(6): the properties of the highest-fluence bursts: pulse ID, estimated redshift, fluence, peak frequency, and isotropic energy.
(7)--(11): the same parameters for the first detected bursts.
Subscripts $\mathrm{m}$ and $\mathrm{f}$ denote parameters corresponding to the highest-fluence and the first detected bursts.
}
\end{deluxetable}

\end{document}